\documentclass[11pt,aps,prd,preprint,onecolumn]{revtex4}
\usepackage{amssymb}
\usepackage{amsmath}
\usepackage{graphics}
\usepackage{latexsym}
\usepackage{array}
\usepackage[activeacute,english]{babel}
\usepackage[latin1]{inputenc}
\usepackage[dvips]{graphicx}
\usepackage{color}

\newcommand{\bn}{\begin{eqnarray}}
\newcommand{\en}{\end{eqnarray}}
\newcommand{\be}{\begin{equation}}
\newcommand{\ee}{\end{equation}}
\newcommand{\no}{\nonumber}
\newcommand{\bc}{\begin{center}}
\newcommand{\ec}{\end{center}}
\newcommand{\la}{\label}
\newcommand{\re}{\ref}

\newcommand{\ci}{\cite}

\begin{document}

\title{\large \bf Electroweak standard model at finite temperature in
presence of a bosonic chemical potential}

\author{C. Pe\~na}
\email{capenac@unal.edu.co}
\author{C. Quimbay}
\email{cjquimbayh@unal.edu.co}
\affiliation{Departamento de F\'{\i}sica, Universidad Nacional de Colombia.\\
Ciudad Universitaria, Bogot\'{a} D.C., Colombia.}

\date{\today}

\begin{abstract}
We study the electroweak standard model at finite temperature in
presence of a bosonic chemical potential associated with the
conserved electromagnetic current. To preserve the thermodynamic
equilibrium of the system, the thermal medium is neutralized by
the introduction of four background charges related to the four
gauge bosons of this model. Using the mean-field approximation, in
the high temperature limit, we find that there exists a difference
between the effective mass of the spatial and temporal components
of the W boson. A W boson condensation induced via
the background charges allows to vanish this difference. \\

\vspace{0.4cm} \noindent {\it{Keywords:}} {Electroweak standard
model, finite temperature, bosonic chemical potential, background
charges, W boson condensation.}

\vspace{0.4cm} \noindent {\it{PACS Codes:}}  11.10.Wx, 11.15.Ex,
14.70.Fm.

\end{abstract}

\maketitle

\section{Introduction}

Since long time ago it is well known that the spontaneous symmetry
breaking for gauge theories at finite temperature in presence of
chemical potentials can be seen as a Bose-Einstein condensation
phenomena \ci{KAP1981, WEL1982}. At this respect, the vacuum
expectation value of the Higgs field $\nu$ at finite temperature
can be seen as a variational parameter of Bose-Einstein
condensation of the Higgs field. The Electroweak Standard Model
(ESM) allows to simultaneously both electroweak phase transition
and Bose-Einstein condensation for the Higgs field if a bosonic
chemical potential $\mu$ related with the conserved
electromagnetic current of the model is considered in the system.
It was found in \ci{KAP1981} that the Critical Temperature ($T_c$)
of the electroweak phase transition increases with $\mu$. This
fact suggest that $\mu$ can be considered as an effective
parameter of spontaneous symmetry breaking. To make the system
neutral, it was necessary in \ci{KAP1981} to add an external
charge background to offset the charge density of the scalar
field.

On the other hand, a zero temperature condensation of SU(2)
vectorial bosons generated by an external lepton density was
proposed in \ci{LIN2}-\ci{KRIVE1980}, for the case in which the
system is electrically neutralized by the inclusion of a external
charge. The generalization at finite temperature of a W boson
condensation was investigated in \ci{PEREZ1}-\ci{KAP1990} for the
case of ESM. This W boson condensation for the ESM can be induced
by the presence of a fermion density in the system. Specifically,
a W boson condensate was investigated by consideration of the
chemical potentials associated with the conserved electromagnetic
and leptonic currents \ci{PEREZ1}. This condensate can also be
investigated by the additional consideration of the chemical
potential associated with the conserved weak neutral current
\ci{FERRER}-\ci{KAP1990}. The existence of the W boson
condensation was studied in references \ci{PEREZ1}-\ci{KAP1990}
through the calculation of the thermodynamical potential, which
leads to the phase diagram of this condensation. It is possible to
see in these references that if the chemical potentials associated
with the conserved leptonic and neutral weak currents vanish, the
chemical potential associated with conserved electromagnetic
current $\mu$ has a zero value in $T=T_c$, being this fact
consistent with the gauge symmetry restoration at high
temperatures. This behavior is not strange because the condensate
considered in these references is neutral and the $T_c$ has not a
dependence over $\mu$. In such references the neutralization of
the thermodynamic system was performed by the inclusion of the
leptonic chemical potential associated with the conserved leptonic
current of the ESM.

The W boson condensation phenomenon has been extensively studied
in the context of neutral many-particle electroweak theory with
the purpose of possible cosmological applications
\ci{LIN2}-\ci{KAP1990}. In this paper we extend these studies
introducing the case of a charged electroweak plasma. The W boson
condensation for the ESM at high temperatures is induced by the
inclusion of background external charges in the thermodynamical
system. The results of this work will have interesting in the
study of charged electroweak plasmas that will be obtained in
future experiments. Specifically the W boson condensation worked
in this paper differs from the one previously studied in the
literature \ci{PEREZ1}-\ci{KAP1990} in the sense that it is
induced but with an external charge instead of a fermion density.
By this reason we do not include in the thermodynamic system the
chemical potentials associated with the conserved leptonic and
neutral weak currents. We only include a bosonic chemical
potential $\mu$ associated with the conserved electromagnetic
current of the ESM. We neutralize the thermal medium by the
introduction of an external charge background which offsets the
charge density of the scalar field, as was performed by Kapusta in
\ci{KAP1981}. Particularly we preserve the thermodynamic
equilibrium of the system by the introduction of four background
charges $j_\nu$, $j^1_\nu$, $j^2_\nu$ and $j^3_\nu$, which are
associated with the four gauge bosons of the ESM \ci{KAP1981,
KAPLIB, LIN1979}. The background charges $j^1_\nu$ and $j^2_\nu$
are associated with the $SU(2)_L$ gauge fields $A^1_\nu$ and
$A^2_\nu$ respectively. It is possible to see that if the
background charges $j^1_\nu$ and $j^2_\nu$ vanish, then the vacuum
average values of the $A^1_\nu$ and $A^2_\nu$ fields vanish too,
i. e. $\langle A^1_\nu\rangle_0=\langle A^2_\nu\rangle_0=0$. We
note that the bosonic chemical potential that we consider here
does not vanish for $T=T_c$, because our interest is focused in
the coexistence of the electroweak broken phase and the Higgs
condensate.

Considering the ESM at finite temperature in presence of $\mu$, in
section II, we calculate the vacuum expectation value of the Higgs
field as a function of $T$ and $\mu$, using the mean-field
approximation in the high temperature limit \ci{KAP1981}. Next we
calculate, in section III, the effective masses of the scalar and
gauge bosons using a generalized gauge. We find that the effective
mass of the spatial component of the W boson has a difference of
$-\mu^2$ respect to its temporal component. In section IV, by
consideration of non-vanishing background charges $j^1_\nu$ and
$j^2_\nu$, we assume that the vacuum average values of the
$A^1_\nu$ and $A^2_\nu$ fields do not vanish. i. e. $\langle
A^1_\nu\rangle_0=\xi_{A^1_\nu}$ and $\langle
A^2_\nu\rangle_0=\xi_{A^2_\nu}$. Because the W field is a
combination of the $A^1_\nu$ and $A^2_\nu$ fields, we obtain a W
boson condensate associated to the spatial component of W boson.
As a consequence of this condensate, the mentioned difference
among the effective masses of the spatial and temporal components
of the W boson vanishes. In section V, by means of the calculation
of the thermodynamic potential, we obtain the electromagnetic
charge density of the system and we calculate the critical
temperature of the W boson condensation. We demonstrate that the W
boson condensate is consistent with the usual condition of
condensation $m^2_W=\mu^2$, where $m_W$ is the effective mass of
the W boson. Finally the conclusions are presented in section VI. \\

\section{Vacuum expectation value of the Higgs field}

The ESM has four conserved currents associated with the four
independent generators of the $SU(2)_L \times U(1)_Y$ gauge group.
The electromagnetic current $J_\nu$ is the only one conventionally
conserved \ci{KAP1981}. We introduce the chemical potential $\mu$
associated with $J_\nu$ in the formalism of the quantum field
theory at finite temperature \ci{KAPLIB} through the partition
function. We initially restrict our interest to the gauge boson
and scalar sectors of the ESM. Performing the functional
integrations over the canonical momentums we obtain that the
effective Lagrangian density ${\cal L}_{eff}$ of the ESM is
\ci{KAP1981} \bn {\cal L}_{eff} &=& [(D^\nu + i\mu Q\delta^{\nu0})
\Phi]^+[(D_\nu + i\mu Q \delta_{\nu0})\Phi] + c^2\Phi^+\Phi -
\lambda(\Phi^+\Phi)^2 \no\\
&&- \frac{1}{4}G^{\mu\nu}G_{\mu\nu}
-\frac{1}{4}\widetilde{F}_a^{\mu\nu}\widetilde{F}^a_{\mu\nu} +
g'B^\nu j_\nu + gA_a^\nu j_{\nu}^a,\la{dle} \en where
$D_\nu=\partial_\nu + igA^a_\nu \tau^a/2+ig'B_\nu/2$ is the $SU(2)_L
\times U(1)_Y$ covariant derivative, $Q=(I+\tau)/2$ is the
electromagnetic charge, $G_{\mu \nu}=\partial_\mu B_\nu -
\partial_\nu B_\mu$ is the $U(1)_Y$ effective strength field tensor,
\be \widetilde{F}_{\mu\nu}^a = \partial_\mu A_\nu^a -\partial_\nu
A_\mu^a - g\epsilon^{abc}(A_\mu^b +
\frac{\mu}{g}\delta_{\mu0}\delta^{b3})(A_\nu^c +
\frac{\mu}{g}\delta_{\nu0}\delta^{c3}) \la{etc} \ee is the
$SU(2)_L$ effective strength field tensor \ci{KAP1981}. We observe
that the form of $\widetilde{F}_{\mu\nu}^a$ implies to shifting
the non-abelian fields by $A_\mu^a \rightarrow  A_\mu^a +
\frac{\mu}{g}\delta_{\mu0}\delta^{a3}$. As was explained in
\ci{KAP1981}, the $\delta_{\mu0}$ arises because there is a
preferred reference frame for the medium, and the $\delta^{a3}$
arises because it is the $A^3_\nu$ field which mixes with $B_\nu$
to form the photon. Because the non-abelian fields have a shift
proportional to $\mu$, then for the case $\mu=0$ the shift
vanishes. Therefore, for this case, (\re{etc}) becomes in the
normal non-abelian strength field tensor and the Lagrangian
density given by (\re{dle}) becomes in the usual for the scalar
and gauge boson sectors of the ESM. The abelian $j_\nu$ and
non-abelian $j_{\nu}^a$ background charges, introduced in
(\re{dle}), are given by \bn
j_\nu=&&j_0\delta_{\nu0},\la{cba} \\
j_\nu^a=&&j_0^3\delta_{\nu0}\delta^{a3}. \la{cbna} \en These background
charges were introduced with the purpose of having a vanishing
electromagnetic charge density, preserving the thermodynamical
equilibrium of the system \ci{KAP1981, FERRER, SAN2003}. Since after
electroweak symmetry breaking the electromagnetism is a $U(1)_1$ symmetry,
it is possible to introduced the electromagnetic background charge
$j_\nu^{em}$ in terms of $j_\nu$ and $j_\nu^3$.

Starting of the effective Lagrangian density (\re{dle}) it is
possible to show that the equation of motion for the scalar field
doublet $\Phi$ is given by  \bn & & \Bigl(\partial^\nu\partial_\nu
+ i\frac{g}{2} (\partial_\nu A_a^\nu)\tau^a +igA_a^\nu
\tau^a\partial_\nu  + i\frac{g'}{2}(\partial_\nu B^\nu) + i g'
B^\nu
\partial_\nu
+ i\mu(I + \tau^3)\partial_0 \no \\
& & - \frac{1}{4}[gA_a^\nu \tau^a + g'B^\nu + \mu \delta^{\nu 0}(I
+ \tau^3)]^2 - c^2\Bigr)\Phi = - 2\lambda\Phi\Phi^+\Phi. \la{emfs}
\en In the ESM, the electroweak symmetry is broken spontaneously
through the Higgs mechanism. As it is shown in \ci{KAP1981}, since
we are considering a system which has an external charge density,
the Higgs mechanism is associated with the existence of a
Bose-Einstein condensate. The implementation of this mechanism is
performed by mean of the following translation of the scalar field
doublet \be \Phi = \frac{1}{\sqrt{2}}\begin{pmatrix} \psi_1 +
i\psi_2\cr \psi_3 + i\psi_4 \end{pmatrix} +
\begin{pmatrix} \xi \cr 0
\end{pmatrix} = \frac{1}{\sqrt{2}} \begin{pmatrix} \nu + \psi_1 +
i\psi_2 \cr \psi_3 + i\psi_4 \end{pmatrix}, \la{thm} \ee where the
vacuum expectation value of the Higgs field $\nu$ has been defined
in terms of the $\xi$ parameter as $\nu = \sqrt{2}\xi$. The $\xi$
parameter represents the infrared stated of the field $\Phi$, and
it can be understood as a variational parameter of the
Bose-Einstein condensation of this scalar field. The choosing of a
vacuum state of the system through (\re{thm}) is consistent with
the fact that this state is energetically preferred for the case
$\mu \neq 0$ \ci{KAP1981}.

Substituting (\re{thm}) into (\re{emfs}), using the mean-field
approximation in which both the field and mixed field fluctuations
vanish \ci{KAP1981}, we obtain that the non-trivial solution for
$\xi$ is \bn \xi^2 &=& \frac{1}{2\lambda}\Bigl[c^2 + \mu^2
-\lambda (3\langle \psi^2_1 \rangle + \langle \psi^2_2 \rangle +
\langle \psi^2_3 \rangle + \langle \psi^2_4 \rangle) + e^2\langle
A^\nu A_\nu \rangle \nonumber\\ && + \frac{1}{4}(g'^2 -
g^2)^2(\frac{e}{gg'})^2 \langle Z^\nu Z_\nu \rangle
 + \frac{1}{2}g^2\langle W^{+\nu} W^-_\nu \rangle \Bigr]\no \\
 &=& \frac{1}{2\lambda}\Bigl[c^2 + \mu^2
- \frac{T^2}{4}\Bigl( \frac{g'^2}{4}+ \frac{3g^2}{4}+2\lambda
\Bigl)\Bigl], \la{ztf}\en where the electromagnetic ($A_\nu$),
neutral electroweak boson ($Z_\nu$) and charged electroweak boson
($W_\nu^\pm$) fields have been defined as usual in terms of the
abelian ($B_\nu$) and non-abelian ($A_\nu^a$) gauge boson fields. We
have used in (\re{ztf}) the definition
$e=gg'/(g'^2+g^2)^\frac{1}{2}$, and the Gibbs averages have been
evaluated in the high temperature limit \ci{LIN1979}, i. e. $
\langle W^{+\nu} W^-_\nu \rangle = \langle Z^\nu Z_\nu \rangle =
\langle A^\nu A_\nu \rangle = \frac{-3T^2}{12}$, $\langle \psi^2_1
\rangle = \langle \psi^2_2 \rangle = \langle \psi^2_3 \rangle =
\langle \psi^2_4 \rangle = \frac{T^2}{12}$.

The $T_c$ of the electroweak phase transition corresponds to the
temperature in which the electroweak symmetry is restored. The value
of $T_c$ is obtained for the temperature in which $\xi$ vanishes,
i.e. $\xi(T_c)=0$. From (\re{ztf}), we obtain that the square of
$T_c$ is given by \ci{KAP1981} \be T^2_c = \frac{4(c^2 +
\mu^2)}{\frac{g'^2}{4} + \frac{3g^2}{4} + 2\lambda}. \la{ctept}\ee
Due $\xi$ is a parameter of Bose-Einstein condensation of the field
$\Phi$, the temperature given by (\re{ctept}) also corresponds to
the critical temperature of the Bose-Einstein condensation of the
same field. The $T_c$ obtained does not contain the fermion
contributions and its value is unknown due $\lambda$ and $\mu$ are
free parameters. We can observe that $\mu$ has the effect to
increase the $T_c$ value. For $\mu = 0$, it is known that $T_c
\approx 200$ GeV.

\section{Effective masses of the bosonic fields}

The equation of motion for the scalar field doublet $\Phi$, given
by (\re{emfs}), leads to the equations of motion for the scalar
fields $\psi_1$, $\psi_2$, $\psi_3$ and $\psi_4$. If we use the
mean-field approximation in these equations of motion, the
following mixing terms vanish
$\bigl\langle[igA^{a\nu}\tau_a\partial^\nu + ig'B^\nu\partial_\nu
+ i\frac{g}{2}\partial_\nu (A^{a\nu}\tau_a) +
i\frac{g'}{2}(\partial^\nu B_\nu)]\phi\bigr\rangle$, $\langle
A^{a\nu} \phi\rangle$, $\langle B^\nu \phi\rangle$, $\langle A^\nu
B_\nu \phi\rangle$. Using the following result that is obtained
from the first-order perturbation theory \ci{KAP1981}
\bn\langle\psi^3_1\rangle+\langle \psi_1\psi^2_2\rangle+\langle
\psi_1\psi^2_3\rangle+\langle
\psi_1\psi^2_4\rangle=\Bigl[3\langle\psi^2_1\rangle
+\langle\psi^2_2\rangle+\langle\psi^2_3\rangle
+\langle\psi^2_4\rangle\Bigr],\en we find that the equations of
motion of the scalars decouple and they can be written as
$(\partial^2+M^2_{(1,2,3,4)eff})\langle \psi_{1,2,3,4}\rangle=0 $,
being $M_{(1,2,3,4)eff}$ the effective masses of the scalar
bosons. We obtain that the square of these effective masses are
\bn M^2_{1eff}&=& - c^2 - \mu^2 +\lambda(3\langle\psi^2_1\rangle +
\langle\psi^2_2\rangle + \langle\psi^2_3\rangle  +
\langle\psi^2_4\rangle) - e^2\langle A^\nu A_\nu\rangle - \no \\&
& \frac{1}{4}(g'^2 - g^2)^2(\frac{e}{gg'})^2\langle Z^\nu
Z_\nu\rangle
- \frac{1}{2}g^2\langle W^{+\nu} W^-_\nu\rangle + 6\lambda \xi^2 \no \\
&=& 4\lambda\xi^2,\la{ems1}\\
M^2_{2eff} &=& - c^2 - \mu^2 + \lambda(3\langle\psi^2_2\rangle +
\langle\psi^2_1\rangle  + \langle\psi^2_3\rangle +
\langle\psi^2_4\rangle)- e^2\langle A^\nu A_\nu \rangle - \no\\
& & \frac{1}{4}(g'^2 - g^2)^2(\frac{e}{gg'})^2 \langle Z^\nu
Z_\nu\rangle - \frac{1}{2}g^2 \langle W^{+\nu} W^-_\nu\rangle +
2\lambda\xi^2,\no \\
&=& 2\lambda(\langle\psi^2_2\rangle - \langle\psi^2_1\rangle ),
\la{ems2}\\
M^2_{3eff}&=&- c^2 + \lambda(3\langle\psi^2_3\rangle +
\langle\psi^2_1\rangle + \langle\psi^2_2\rangle +
\langle\psi^2_4\rangle) - \frac{1}{2}g^2\langle W^{+\nu}W^-_\nu
\rangle -\no\\
& & \frac{1}{4}(\frac{gg'}{e})^2\langle Z^\nu
Z_\nu\rangle  + 2\lambda\xi^2,\no \\
&=& 2\lambda(\langle\psi^2_3\rangle - \langle\psi^2_1\rangle ),\la{ems3}\\
M^2_{4eff}&=& - c^2 + \lambda(3\langle\psi^2_4\rangle +
\langle\psi^2_1\rangle + \langle\psi^2_2\rangle +
\langle\psi^2_3\rangle) - \frac{1}{2}g^2\langle
W^{+\nu}W^-_\nu\rangle -\no\\
& & \frac{1}{4}(\frac{gg'}{e})^2\langle Z^\nu Z_\nu\rangle  +
2\lambda\xi^2,\no \\
&=& 2\lambda(\langle\psi^2_4\rangle - \langle\psi^2_1\rangle
).\la{ems4} \en We observe that in the high temperature limit
$M^2_{1eff}=2\Bigl[c^2 + \mu^2 - \frac{T^2}{4}\Bigl( \frac{g'^2}{4}+
\frac{3g^2}{4}+2\lambda \Bigl)\Bigl]$ and
$M^2_{2eff}=M^2_{3eff}=M^2_{4eff}=0$. The effective masses obtained
for the scalar particles do not include the effects of performing
the gauge fixing. In the next we will use the renormalizable
$R_\rho$ gauge \ci{KAP1990}.

The effective Lagrangian density (\re{dle}) leads to the effective
equations of motion for the non-physical gauge bosons $B^\nu$ and
$A^{a\nu}$. It is possible to prove that the effective equation
for the abelian gauge boson $B^\nu$ is \be \la{eeabe}
 \partial^\mu G_{\mu\nu} =
-i\frac{g'}{2}\Bigl[[(D_\nu + i\mu Q \delta_{\nu0})\phi]^+\phi -
\phi^+(D_\nu + i\mu Q \delta_{\nu0})\phi\Bigr] - g'j_\nu, \ee
while the ones for the non-abelian gauge bosons $A^{a\nu}$ are \bn
\la{eenoa}
\partial^\mu\widetilde{F}^a_{\mu\nu} &=& -i\frac{g}{2}
\Bigl\{\bigl[(D_\nu + i\mu Q \delta_{\nu0})\phi\bigr]^+\tau^a\phi
- \phi^+\tau^a[(D_\nu + i\mu Q \delta_{\nu 0})\phi]\Bigr\}
                  \nonumber\\
& & - \epsilon^{abc}\widetilde{F}^b_{\mu\nu}(gA^{c\mu} +
\mu\delta^{\mu0}\delta^{c3}) - gj^a_\nu, \en being $a=1,2,3$.
After substituting (\re{thm}) into (\re{eeabe}), taking the
statistical average, without considering fluctuations of the
abelian gauge field, i. e. $<B_\nu>=0$, we obtain that the
background abelian charge $j_\nu$ is given by \be\la{bac} j_\nu =
\frac{-i}{2}\Bigl\langle(\widetilde{D}_\nu\phi)^+\phi - \phi^+
(\widetilde{D}_\nu\phi)\Bigr\rangle, \ee where $\widetilde{D}^\nu
= D^\nu + i\mu Q\delta^{\nu 0}$. In the mean-field approximation,
the mixed field terms of (\re{bac}) can be forgotten and we obtain
\be \la{bacuv} j_\nu = -\mu\delta_{\nu 0} \frac{1}{2}\Bigr(2\xi^2
+ \langle\psi^2_1\rangle + \langle\psi^2_2\rangle\Bigl), \ee being
$j_\nu$ proportional to $\mu$. It is clear that for a vanishing
chemical potential this background abelian charge vanishes. In a
similar way, starting from (\re{eenoa}), without considering
fluctuations as $\langle\partial^\nu A^a_\nu\rangle_0=0$, we
obtain that the background non-abelian charges $j^a_\nu$ have the
form \bn\la{nac} j^a_\nu =
\frac{-i}{2}\Bigl[\bigl\langle(\widetilde{D}_\nu\phi)^+
\tau^a\phi\rangle - \langle\phi^+
\tau^a(\widetilde{D}_\nu\phi)\bigr\rangle\Bigr] -
\frac{1}{g}\epsilon^{abc}\bigl\langle\widetilde{F}^b_{\mu\nu}(gA^{c\mu
} + \mu\delta^{\mu 0}\delta^{c3})\bigr\rangle. \en After
substituting (\re{thm}) into (\re{nac}), we obtain the three
following background charges: \bn \la{thnac} j^1_\nu &=&
-\frac{\mu^2}{g}\bigl[\langle A^1_0\rangle_0\delta_{\nu 0}-
\langle A^1_\nu\rangle_0\bigr],\la{m128.01}\\
j^2_\nu &=& -\frac{\mu^2}{g}\bigl[\langle
A^2_0\rangle_0\delta_{\nu 0}- \langle
A^2_\nu\rangle_0\bigr]\la{m128.02},\\
j^3_\nu &=& - \frac{1}{2}\mu\delta_{\nu 0}\bigl[2\xi^2 +
\langle\psi^2_1\rangle + \langle\psi^2_2\rangle\bigr]
                             \nonumber\\
&&+\mu\biggl[\Bigl(\bigl\langle(A^1_\nu)^2\bigr\rangle+
\bigl\langle(A^2_\nu)^2\bigr\rangle\Bigr) \delta_{\nu
0}-\Bigl(\bigl\langle A^1_\nu A^{10}\bigr \rangle+\bigl\langle
A^2_\nu A^{20}\bigr\rangle\Bigr)\biggr],\la{m128.1} \en where we
note that these three background charges are also proportional to
$\mu$. As we have defined $j^a_\nu=j^3_0\delta^{a3}\delta_{\nu 0}$
in (\re{cbna}), this means that $j^1_\nu=j^2_\nu=0$.  According
with (\re{m128.01}) and (\re{m128.02}), it is necessary by
consistency that $\langle A^1_i\rangle_0=\langle
A^2_i\rangle_0=0$. It is clear that the background charges $j_\nu$
y $j^3_\nu$ are not conserved because they have a dependence over
the temperature. Additionally, because there are not vacuum
fluctuations for the non-abelian gauge fields $A^1_\nu$ y
$A^2_\nu$, then there not exists a W boson condensation due
$\langle W^\pm_\nu\rangle_0=\frac{1}{\sqrt{2}}(\langle A^1_\nu
\rangle_0\mp i\langle A^2_\nu\rangle_0)=0$.

Using the mean-field approximation, in which the mixing terms that
appears in the equation of motion (\ref{eeabe}) and (\ref{eenoa})
vanish, we obtain that the effective masses of the non-physical
gauge bosons are given by
\begin{eqnarray}
M^2_{eff B_\nu} &=& \frac{g'^2}{4}\bigl(2\xi^2 +
\langle\psi^2_1\rangle + \langle\psi^2_2\rangle +
\langle\psi^2_3\rangle + \langle\psi^2_4\rangle\bigr),\la{m193.1}\\
M^2_{eff A^1_\nu} &=& M^2_{eff A^2_\nu} = \frac{g^2}{4}(2\xi^2 +
\langle\psi^2_1\rangle + \langle\psi^2_2\rangle +
\langle\psi^2_3\rangle +
\langle\psi^2_4\rangle) + q_\nu,\la{m170}\\
M^2_{eff A^3_\nu} &=& \frac{g^2}{4}(2\xi^2 + \langle\psi^2_1\rangle
+ \langle\psi^2_2\rangle + \langle\psi^2_3\rangle +
\langle\psi^2_4\rangle),\la{m193.2}
\end{eqnarray}
being
\begin{equation}\la{m13} q_\nu = \left\{
\begin{array}{ll}
 0, & \textrm{if $\nu=0$.}\\
-\mu^2, & \textrm{if $\nu=i$.}
     \end{array} \right.
\end{equation}
Because the W boson is defined as
$W^\pm_\nu=\frac{1}{\sqrt{2}}(A^1_\nu \mp i A^2_\nu)$, we observe
that the effective mass of the spatial component of the W boson
has a difference of $-\mu^2$ respect to its temporal component.

The quantity $\langle\psi^2_1\rangle + \langle\psi^2_2\rangle +
\langle\psi^2_3\rangle + \langle\psi^2_4\rangle$, which appears in
(\re{m193.1}), (\re{m170}) and (\re{m193.2}), has its origin in the
term $\Bigl(\psi^2_1+\psi^2_2+\psi^2_3+\psi^2_4\Bigr)
\Bigl[(A^1_\nu)^2+(A^2_\nu)^2+(A^3_\nu)^2+(B_\nu)^2\Bigr]$ of the
Lagrangian density (\re{dle}) and contributes also to the effective
masses of the scalar bosons. With the purpose to include only the
terms that contribute to the effective masses of the bosons, we
obtain that the effective action $S_E$, in the imaginary time
formalism, is given by
\begin{eqnarray}\la{m212}
S_E &=& {\int d^4x}\biggl[ - \frac{1}{2}\psi_1(\partial^2 +
M^2_{1eff})\psi_1 - \frac{1}{2}\psi_2(\partial^2 +
M^2_{2eff})\psi_2 - \frac{1}{2}\psi_3(\partial^2 +
M^2_{3eff})\psi_3
                \nonumber\\
&-& \frac{1}{2}\psi_4(\partial^2 + M^2_{4eff})\psi_4 + \xi^2(\mu^2
+ c^2 - \lambda\xi^2) + \frac{1}{2}A^{3\nu}\Bigl[\bigl(\partial^2
+
\frac{g^2}{2}\xi^2\bigr)g_{\mu\nu}-\partial_\nu\partial_\mu\Bigr]A^{3\mu}
              \nonumber\\
&+&\frac{1}{2}A^{1\nu}\Bigl[\bigl(\partial^2 + \frac{g^2}{2}\xi^2
+ q^\nu\bigr)g_{\mu\nu}-\partial_\nu\partial_\mu\Bigr]A^{1\mu}
             \nonumber\\
&+& \frac{1}{2}A^{2\nu}\Bigl[\bigl(\partial^2 + \frac{g^2}{2}\xi^2
+ q^\nu\bigr)g_{\mu\nu}-\partial_\nu\partial_\mu\Bigr]A^{2\mu}
                 \nonumber\\
&+& \frac{1}{2}B^\nu[\bigl(\partial^2 +
\frac{g'^2}{2}\xi^2\bigr)-\partial_\nu\partial_\mu]B^\nu +
\frac{1}{2}gA^{3\nu}\sqrt{2}\xi\partial_\nu\psi_2 -
\frac{1}{2}gA^{2\nu}\sqrt{2}\xi\partial_\nu\psi_3
                  \nonumber\\
&+& \frac{1}{2}gA^{1\nu}\sqrt{2}\xi\partial_\nu\psi_4  +
\frac{1}{2}g'B^\nu\sqrt{2}\xi\partial_\nu\psi_2\biggr],
\end{eqnarray}
where we have depressed the terms of ${\cal L}_{eff}$ which do not
contribute to the equations of motion and we have only preserved the
terms which can be canceled with the gauge fixing.

In this paper we work with the renormalizable $R_\rho$ gauge
\ci{KAP1990}. For the $U(1)_Y$ abelian part, the gauge fixing
function is given by
\begin{eqnarray}\la{m213}
F&=&\partial^\nu B_\nu -\frac{g'}{2}\sqrt{2}\xi\rho\psi_2 - f(x,\tau),\\
\frac{\partial F}{\partial \alpha} &=& -\partial^2 -
2(\frac{g'}{2})^2\rho\xi^2,
\end{eqnarray}
being $\alpha$ the abelian gauge transformation phase given by
$B_\nu\rightarrow B'_\nu=B_\nu-\partial_\nu\alpha(x,\tau)$, while
for the $SU(2)_L$ non-abelian gauge part, the gauge fixing function
is given by:
\begin{eqnarray}\la{m214}
F^1&=&\partial^\nu A^1_\nu -\frac{g}{2}\sqrt{2}\xi\rho\psi_4
- f^1(x,\tau),\\
F^2&=&\partial^\nu A^2_\nu +\frac{g}{2}\sqrt{2}\xi\rho\psi_3
- f^2(x,\tau),\\
F^3&=&\partial^\nu A^3_\nu -\frac{g}{2}\sqrt{2}\xi\rho\psi_2 -
f^3(x,\tau).
\end{eqnarray}
Because $f^a(x,\tau)$ labels the three gauge arbitrary functions
and $\alpha^a(x,\tau)$ labels the non-abelian gauge transformation
phases
\begin{eqnarray}\la{m216}
A^c_\nu \rightarrow A'^c_\nu = A^c_\nu +
g\epsilon^{abc}A^a_\nu\alpha^b(x,\tau) -
\partial_\nu\alpha^c(x,\tau),\end{eqnarray} then
\begin{eqnarray}\la{m217}
\frac{\partial F^i}{\partial \alpha^i} &=& -\partial^2 -
(\frac{g^2}{2})\rho\xi^2,\,\,\,\textrm{for $i=1,2,3.$} \la{m217.3}
\end{eqnarray}
We observe that our gauge fixing functions $F$ and $F^a$ do not have
a dependence over $\mu$ and this fact will allow to obtain a gauge
invariant thermodynamical potential in the high temperature limit
which does not depend over the $\rho$ parameter.\\

The effective partition function $Z_E$ that we obtain working in
the generalized gauge fixing and in the imaginary time formalism
is \bn\la{m218} Z_E &=&
\biggl(det\Bigl[\frac{\partial^2}{\partial\tau^2}+\nabla^2-
m^2_{C^a}\Bigr]\biggr)^3
\biggl(det\Bigl[\frac{\partial^2}{\partial\tau^2}+\nabla^2 -
m^2_C\Bigr]\biggr)
                   \nonumber\\
& &\times\int\limits_\textrm{\scriptsize\it
periodic}{D[\psi_1]D[\psi_2]D[\psi_3]D[\psi_4]
D[A^1_\nu]D[A^2_\nu]D[A^3_\nu]D[B_\nu]}
                \nonumber\\
& & \times\int{D[f^1]D[f^2]D[f^3]
D[F]\delta(F^1)\delta(F^2)\delta(F^3)\delta(F)}
\nonumber\\
& & \times\exp \biggl\{S_E-\int^\beta_0 d\tau\int_V d^3x
\frac{1}{2\rho}(f^a)^2 +\frac{1}{2\rho}(f)^2\biggr\},\en where the
two determinants which appear in this partition function are the
Fadev-Popov's determinants and they contain the contribution of the
ghost fields $C^a$ and $C$, whose masses are $m^2_{C^a}=
2\bigl(\frac{g}{2}\bigr)^2\rho\xi^2$ and $\ m^2_C=
2\bigl(\frac{g'}{2}\bigr)^2\rho\xi^2$, respectively. Now it is
possible to identify in the argument of the exponential the bosonic
effective masses. The squares of the effective masses of the scalar
bosons are
\bn m^2_{1}(\xi)&=&M^2_{1eff},\la{m1}\\
m^2_{2}(\xi)&=&M^2_{2eff}+ \frac{g^2}{2}\rho\xi^2 + \frac{g'^2}{2}
\rho\xi^2,\la{m2}\\
m^2_{3}(\xi)&=&M^2_{3eff}+ \frac{g^2}{2}\rho\xi^2,\la{m3}\\
m^2_{4}(\xi)&=&M^2_{4eff}+ \frac{g^2}{2}\rho\xi^2,\la{m4}\en where
the effective masses $M^2_{1eff}$, $M^2_{2eff}$, $M^2_{3eff}$ and
$M^2_{4eff}$ are given by (\re{ems1}), (\re{ems2}), (\re{ems3}) and
(\re{ems4}), respectively. In the high temperature limit, the
effective masses of the Higgs and Goldstone bosons are:
\bn m^2_{H}(\xi)&=&2\lambda (2\xi^2),\la{mh1}\\
m^2_{\phi^0}(\xi)&=&\frac{g^2+g'^2}{4}\rho (2\xi^2),\la{mph0}\\
m^2_{\phi^{\pm}}(\xi)&=&\frac{g^2}{4}\rho (2\xi^2),\la{mphpm}\en
being $2\xi^2=\frac{1}{\lambda}\Bigl[c^2 + \mu^2 -
\frac{T^2}{4}\Bigl( \frac{g'^2}{4}+ \frac{3g^2}{4}+2\lambda
\Bigl)\Bigl]$. We note that the effective masses of the Goldstone
bosons depend on the gauge parameter $\rho$ and therefore the
Goldstone bosons are not part of the physical particle spectrum. On
the other hand, the squares of the effective masses of the
non-physical gauge bosons are given by
\begin{eqnarray}
m^2_{B_\nu}(\xi) &=& \frac{g'^2}{4}\bigl(2\xi^2\bigr),\la{m170.1}\\
m^2_{A^1_\nu}(\xi,\mu) &=& m^2_{A^2_\nu}(\xi,\mu) = \frac{g^2}{4}
(2\xi^2)+q_\nu,\la{m193.11}\\
m^2_{A^3_\nu}(\xi) &=& \frac{g^2}{4}(2\xi^2).\la{m193.21}
\end{eqnarray} Starting from (\re{m170.1}), (\re{m193.11}) and
(\re{m193.21}), we can write the squares of the effective masses of
the physical gauge bosons as
\begin{eqnarray}
m^2_{A_\nu}(\xi)&=&\bigl(\frac{g'^2}{g^4 -
g'^4}\bigr)\bigl(\frac{g}{e}\bigr)^2[g^2m^2_{B_\nu}
- g'^2m^2_{A^3_\nu}] \nonumber\\&=& 0\la{m194.1},\\
m^2_{Z_\nu}(\xi) &=& \bigl(\frac{g^2}{g'^4 -
g^4}\bigr)\bigl(\frac{g'}{e}\bigr)^2[g'^2m^2_{B_\nu} - g^2m^2_{
A^3_\nu}]\nonumber\\
&=&\frac{g^2+g'^2}{4\lambda}\bigr[c^2+\mu^2-\frac{T^2}{4}
\bigr(\frac{g'^2}{4} +\frac{3g^2}{4} +2\lambda \bigr)\bigr],\la{m194.2}\\
m^2_{W^\pm_\nu}(\xi,\mu) &=& m^2_{A^1_\nu}(\xi,\nu)=
m^2_{A^2_\nu}(\xi,\mu)\no \\ &=&
\frac{g^2}{4\lambda}\bigr[c^2+\mu^2-\frac{T^2}{4}
\bigr(\frac{g'^2}{4} +\frac{3g^2}{4} +2\lambda
\bigr)\bigr]+q_\nu.\la{m194.3}
\end{eqnarray}
The effective masses given by (\re{m194.2}) and (\re{m194.3})
allow to the physical masses of the electroweak gauge bosons
taking $T=0$ and $\mu=0$. For the case $T \neq 0$ and $\mu=0$,
these masses are in agreement with the effective masses given in
\ci{GAV1999}. We note that the effective mass of the W boson is a
function of the amount $q_\nu$, this implies that the effective
mass of the spatial component of the W boson has a difference of
$-\mu^2$ respect to its temporal component. This behavior is in
agreement with the observation realized by Kapusta in \ci{KAP1990}
when he mentioned that the transverse W's have associated a
chemical potential $\mu$ whereas the longitudinal W's does not
have one. This strange behavior is contrary to what happens in the
non-gauge theories where all three spin states of a massive spin-1
bosons have associated a common chemical potential \ci{KAP1990}.

For high temperatures above the $T_c$ of the electroweak phase
transition, i. e. $T\geq T_C$, we observe from (\re{m194.2}) that
the effective mass of the Z-gauge boson vanishes because
$\xi^2=0$, while from (\re{m194.3}) the effective mass for the W
boson non-vanishes, and it takes the value
$m_{W^\pm_\nu}(\xi,\mu)=q_\nu$.

\section{Induction of a W boson condensate}

In the effective Lagrangian density of the ESM, given by (\re{dle}),
we have introduced the abelian $j_\mu$ and non-abelian $j_\mu^3$
background charges with the purpose to have a vanishing
electromagnetic charge density, and so, to preserve the
thermodynamic equilibrium of the system. It is important to note
that a non-abelian gauge theory with external charges requires
a special quantization scheme as was studied in \ci{FER1987}. However
we can introduce the non-abelian background charges $j_\mu^1$
and $j_\mu^2$, in such a way that the thermodynamic equilibrium of
the system is not affected. These non-abelian charges should be only
dynamical, and satisfying the neutrality condition, with the purpose
to preserve the gauge symmetry of the theory \ci{FER1987}. The fact of
considering $j_\mu^1 \ne 0$ and $j_\mu^2 \ne 0$ implies from
(\re{m128.01}) and (\re{m128.02}) that $\langle A^1_i\rangle_0 \ne 0$
and $\langle A^2_i\rangle_0 \ne 0$. We parameterize the non-vanishing
vacuum average values of the spatial components of the $A^1_\nu$ and
$A^2_\nu$ fields as \bn \langle
A^1_i\rangle_0&=&\xi_{A^1_i}(x),\la{c1}\\
\langle A^2_i\rangle_0&=&\xi_{A^2_i}(x),\la{c2}\en where they
acquire non-homogeneous classical components. The non-homogeneity
is shown by the dependence of $\xi_{A^1_i}(x)$ and
$\xi_{A^2_i}(x)$ on the space-time, consequently the later means
that the vacuum average value of the spatial components of the W
boson fields are non-homogeneous, and it is given by \bn\langle
W^\pm_i\rangle_0(x)=\xi_{W^\pm_i}(x)=\frac{1}{\sqrt{2}}\bigl
(\xi_{A^1_i}(x)\mp i \xi_{A^2_i}(x)\bigr).\la{c3} \en This fact
implies that we have obtained a W boson condensation associated
with the spatial component of the W boson for a system in which
was not included the chemical potentials associated with the
conserved leptonic and neutral weak currents. The background
charges (\re{m128.01}) and (\re{m128.02}) are written as \bn g
j^1_\nu &=&-\mu^2\bigl[\langle A^1_0\rangle_0\delta_{\nu 0}-
\langle
A^1_\nu\rangle_0\bigr]=-\mu^2\bigr[\xi_{A^1_0}(x)\delta_{\nu 0
}-\xi_{A^1_\nu}(x)\bigr],\la{c4.1}\\
g j^2_\nu &=&-\mu^2\bigl[\langle A^2_0\rangle_0\delta_{\nu 0}-
\langle
A^2_\nu\rangle_0\bigr]=-\mu^2\bigr[\xi_{A^2_0}(x)\delta_{\nu 0
}-\xi_{A^2_\nu}(x)\bigr].\la{c4.2} \en We observe that the time
components of $j^1_\nu$ and $j^2_\nu$ vanish, while the spatial
components of these charges do not. This fact means that the
non-abelian background charges given by (\re{c4.1}) and
(\re{c4.2}) are \bn g j^1_\nu
&=&\mu^2\xi_{A^1_\nu}(x),\la{c4.11}\\
 g j^2_\nu &=&\mu^2\xi_{A^2_\nu}(x),\la{c4.22}
 \en where is clear that $\xi_{A^1_0}(x)=\xi_{A^2_0}(x)=0$.

If we use the mean-field approximation in the equation of motion
(\re{eenoa}), modified by the introduction of the two news
background charges, we obtain \bn
\Bigl((\partial^2g^\mu_\nu-\partial^\mu\partial_\nu)
+\frac{g^2}{4}\bigl(2\xi^2_1 +
\langle\psi^2_1\rangle+\langle\psi^2_2\rangle+\langle\psi^2_3\rangle
+\langle\psi^2_4\rangle\bigr)+ q\Bigr)\langle A^1_\mu\rangle=-g
j^1_\nu,\la{c5}\\
\Bigl((\partial^2g^\mu_\nu-\partial^\mu\partial_\nu)
+\frac{g^2}{4}\bigl(2\xi^2_1 +
\langle\psi^2_1\rangle+\langle\psi^2_2\rangle+\langle\psi^2_3\rangle
+\langle\psi^2_4\rangle\bigr)+ q\Bigr)\langle
A^2_\mu\rangle=-gj^2_\nu,\la{c6}\en where the square of the vacuum
expectation value of the Higgs field $\xi_1$, for this case, is
\bn \xi^2_1=\xi^2+\frac{1}{2\lambda}
\Bigl(\frac{g^2}{4}(\xi_{W^+_\nu})^2
+\frac{g^2}{4}(\xi_{W^-_\nu})^2\Bigr),\la{c7}\en being
$(\xi_{W^+_\nu})$ and $(\xi_{W^-_\nu})$ the parameters of W boson
condensation.

The effective masses of the $A^1_\nu$ and $A^2_\nu$ fields can be
obtained from the equations of motion of these fields in the
vacuum state. If we substitute (\re{c4.11}) and (\re{c4.22}) into
(\re{c5}) and (\re{c6}), respectively, and remembering that
$\langle A^{1,2}_\nu\rangle_0=\xi_{A^{1,2}_\nu}(x)$, we obtain.
\bn
\Bigl((\partial^2g^\mu_\nu-\partial^\mu\partial_\nu)+\frac{g^2}{4}
(2\xi^2_1 +\langle\psi^2_1\rangle+\langle\psi^2_2\rangle
+\langle\psi^2_3\rangle+\langle\psi^2_4\rangle)
\Bigr)\xi_{A^{1,2}_\mu}(x)=0,\la{p1}\en where is clear the
cancelation of the term $q$. For an excited state it is also
possible to find the effective masses of the non-abelian gauge
fields by the consideration of a small variation of the vacuum
state as $\langle
A^{1,2}_\mu\rangle\rightarrow\xi_{A^{1,2}_\mu}(X) +
\delta\xi_{A^{1,2}_\mu}(X)$. Applying this result into (\re{c5})
and (\re{c6}), we obtain \bn
\Bigl((\partial^2g^\mu_\nu-\partial^\mu\partial_\nu)+\frac{g^2}{4}
(2\xi^2_1 +\langle\psi^2_1\rangle+\langle\psi^2_2\rangle
+\langle\psi^2_3\rangle+\langle\psi^2_4\rangle+q)
\Bigr)\delta\xi_{A^{1,2}_\mu}(x)=0,\en due there not exist a W
boson condensate in an excited state, i. e.
$\xi_{A^{1,2}_\mu}(x)=0$. So, the effective masses of the
non-abelian gauge bosons are \bn M^2_{effA^{1,2}_\nu}=\left\{
\begin{array}{ll}
 \frac{g^2}{4}(2\xi^2_1
+\langle\psi^2_1\rangle+\langle\psi^2_2\rangle
+\langle\psi^2_3\rangle+\langle\psi^2_4\rangle+q), &
\textrm{(Excited state),}\\ \frac{g^2}{4}(2\xi^2_1
+\langle\psi^2_1\rangle+\langle\psi^2_2\rangle
+\langle\psi^2_3\rangle+\langle\psi^2_4\rangle), & \textrm{(Vacuum
state)},
     \end{array} \right.\en
being clear that the term $q$ vanish in the vacuum state of the
$A^1_\nu$ and $A^2_\nu$ fields. This vacuum state corresponds to
the phase of W boson condensation. In the condensation phase the
effective mass of the spatial components of the W boson is equal
to its temporal component. The difference among the effective
masses of the spatial and temporal components of the W boson has
vanished as a consequence of the introduction of the non-abelian
background charges $j^{1}_\nu$ and $j^{2}_\nu$ given by
(\re{c4.11}) and (\re{c4.22}), respectively.

\section{Critical temperature of the W boson condensation}

The critical temperature of the W boson condensation $T_w$ can be
obtained from (\re{ztf}) and (\re{c7}), for the case in which
$(\xi_{W^\pm_\nu})^2=0$. This critical temperature is given by \bn
T^2_{w}=\frac{4\Bigl(c^2 + \mu^2-
2\lambda\xi^2_1(T_{w})\Bigr)}{\frac{1}{4}g'^2 + \frac{3}{4}g^2 +
2\lambda},\la{c7.1}\en where $\xi^2_1$ is evaluated in $T=T_w$.
The expression (\re{c7.1}) is not useful to calculate $T_{w}$
because $\xi^2_1(T_{w})$ is unknown. To know $T_{w}$ it is first
necessary to calculate the thermodynamic potential $\Omega_f$,
using the mean-field approximation, in the high temperature limit.

To calculate $\Omega_f$, by convenience, we initially calculate
the thermodynamic potential $\Omega_0$ that we obtain for the case
in which the background charges $j_\mu$ and $j^3_\mu$ are not
included. The effective partition function $Z^0_E$
for this case is given by \bn\la{c10} Z^0_E &=&
\biggl(det\Bigl[\frac{\partial^2}{\partial\tau^2}+\nabla^2 -
m^2_{C^a}\Bigr]\biggr)^3
\biggl(det\Bigl[\frac{\partial^2}{\partial\tau^2}+\nabla^2 -
m^2_C\Bigr]\biggr)
                   \nonumber\\
& &\times\int\limits_\textrm{\scriptsize\it
periodic}{D[\psi_1]D[\psi_2]D[\psi_3]D[\psi_4]
D[\delta\xi_{A^{1}_\nu}]D[\delta\xi_{A^{2}_\nu}]D[A^3_\nu]D[B^\nu]}
                \nonumber\\
& & \times\int{D[f^1]D[f^2]D[f^3]
D[F]\delta(F^1)\delta(F^2)\delta(F^3)\delta(F)}\, \exp [A],\en
where the effective action $A$ is \bn\la{c11} A&=&{\int^\beta_0
d\tau\int_Vd^3x\,}\Biggl\{
\frac{1}{2}\psi_1\Bigl(\frac{\partial^2}{\partial\tau^2}+\nabla^2-
m^2_1(\xi_1)\Bigr)\psi_1
+\frac{1}{2}\psi_2\Bigl(\frac{\partial^2}{\partial\tau^2}+\nabla^2-
m^2_2(\xi_1)\Bigr)\psi_2
              \nonumber\\
&&+\frac{1}{2}\psi_3\Bigl(\frac{\partial^2}{\partial\tau^2}+\nabla^2-
m^2_3(\xi_1)\Bigr)\psi_3
+\frac{1}{2}\psi_4\Bigl(\frac{\partial^2}{\partial\tau^2}+\nabla^2-
m^2_4(\xi_1)\Bigr)\psi_4
               \nonumber\\
&&+\frac{1}{2}\xi_{A^{1\mu}}(x,\tau)\Bigl(\bigl(\frac{\partial^2}
{\partial\tau^2}+\nabla^2-\frac{g^2}{2}\xi^2_1\bigr)\delta_{\mu\nu}
+\frac{1-\rho}{\rho}\partial_\mu\partial_\nu\Bigr)\xi_{A^{1\nu}}(x,\tau)
               \nonumber\\
&&+\frac{1}{2}\xi_{A^{2\mu}}(x,\tau)\Bigl(\bigl(\frac{\partial^2}
{\partial\tau^2}+\nabla^2-\frac{g^2}{2}\xi^2_1\bigr)\delta_{\mu\nu}+
\frac{1-\rho}{\rho}\partial_\mu\partial_\nu\Bigr)\xi_{A^{2\nu}}(x,\tau)
               \nonumber\\
&&+\frac{1}{2}A^{3\mu}\Bigl(\bigl(\frac{\partial^2}{\partial\tau^2}+
\nabla^2-\frac{g^2}{2}\xi^2_1\bigr)\delta_{\mu\nu}+\frac{1-\rho}{\rho}
\partial_\mu\partial_\nu\Bigr)A^\nu_3
                \nonumber\\\
&&+\frac{1}{2}B^\mu\Bigl(\bigr(\frac{\partial^2}{\partial\tau^2}+
\nabla^2-
\frac{g'^2}{2}\xi^2_1\bigr)\delta_{\mu\nu}+\frac{1-\rho}{\rho}
\partial_\mu\partial_\nu\Bigr)B^\nu
                \nonumber\\
&&+\xi^2_1\bigl(\mu^2+c^2-\lambda\xi^2_1\bigr)+\frac{g^2}{4}\xi^2_1
\Bigl[(\xi_{W^+_\nu})^2+(\xi_{W^-_\nu})^2\Bigr]
           \no\\
&&+\frac{1}{2}\mu^2\Bigl[(\xi_{W^+_\nu})^2+(\xi_{W^-_\nu})^2\Bigr]
\Biggr\}.\en After performing the functional integrals over $\ln
Z^0_E$, using the high temperature limit, we obtain that the
thermodynamic potential $\Omega_0$ is \bn\la{c12} \Omega_0 =
-\frac{\ln Z^0_E}{\beta V}&=& -\xi^2_1\Bigl[\mu^2
+c^2-\Bigl(\frac{3g^2}{4}+\frac{g'^2}{4}+2\lambda\Bigr)\frac{T^2}{4}
\Bigr] -(2\mu^2+4c^2)\frac{T^2}{24}
                         \nonumber\\
&&+\frac{T^4}{24}\Bigl(\frac{3g^2}{4}+\frac{g'^2}{4}+2\lambda\Bigr)
-\frac{2\pi^2}{15}T^4
               \nonumber\\
&&+\lambda\xi^4_1-\frac{g^2}{4}\xi^2_1\Bigl((\xi_{W^+_\nu})^2+
(\xi_{W^-_\nu})^2\Bigr).\en We observe that $\Omega_0$ does not
dependent from the gauge parameter $\rho$. For the case $\mu=0$, we
obtain that (\re{c12}) is in agreement with the thermodynamic
potential known in the literature \ci{KAPLIB}.

Starting from (\re{c12}) we find that the electromagnetic charge
density $\sigma_0$ for the non-neutralized system is \bn\la{c13}
\sigma_0=-\frac{\partial\Omega_0}{\partial\mu} =\frac{\mu T^2
}{6}+2\mu\xi^2_1.\en

If we include the background charges $j_\mu$ and $j^3_\mu$, the
system has now a vanishing electromagnetic charge density. The
electromagnetic charge density $\sigma$ for the neutralized system
is given by \bn\la{c15} \sigma=\frac{1}{\beta V }\frac{\partial\ln
Z}{\partial\mu}=\frac{1}{\beta V }\frac{1}{Z}\frac{\partial
Z}{\partial\mu},\en where $Z$ is the partition function given by
(\re{m218}) modified by the inclusion of the non-abelian
background charges $j^1_\mu$ and $j^2_\mu$. Using the mean-field
approximation, in the high temperature limit, we obtain that
\bn\la{d1} \sigma=\sigma_0+\sigma_1, \en where $\sigma_0$ is given
by (\re{c13}) and $\sigma_1$ is \bn \sigma_1 &=&
-\frac{1}{4}\frac{\partial}{\partial\mu} \langle
\widetilde{F}^a_{\mu\nu}\widetilde{F}^{\mu\nu}_a\rangle+
\frac{\partial}{\partial\mu}\langle g'B^0J_0+gA_3^0J_0^3\rangle
\nonumber\\ &=&\frac{3\mu
T^2}{6}+\frac{\partial}{\partial\mu}\langle
g'B^0J_0+gA_3^0J_0^3\rangle. \en Because the system has been
neutralized, then the electromagnetic charge density vanishes, i.
e. $\sigma=\sigma_0+\sigma_1=0$, and then \bn\la{d2} \langle
g'B^0J_0+gA_3^0J_0^3\rangle=-\frac{1}{3}\mu^2 T^2-\int
2\mu\xi^2_1\,d\mu +f(T),\en where $f(T)$ is an arbitrary function
over temperature, that we choose as zero. For $T=T_c$, we obtain
that the background charges are \bn j_0\biggl|_{T=T_c}=-\frac{\mu
T^2_c}{12},\la{d3.1}\\
j^3_0\biggl|_{T=T_c}=-\frac{7\mu T^2_c}{12}.\la{d3.2}\en
Substituting (\re{d3.1}) and (\re{d3.2}) into (\re{d2}), and
assuming a non-vanishing chemical potential we obtain that
\bn\la{d4} (g'\langle B^0\rangle+7g\langle A^0_3\rangle)=4\mu. \en
We observe, for $T=T_c$, that a possible solution of the last
equation is $\langle B^0\rangle=\frac{\mu}{2g'}$ and $\langle
A^0_3\rangle=\frac{\mu}{2g}$. This means that any value of $\mu$
satisfies $\sigma=0$, and then $\mu$ can describes a phase
transition.

The thermodynamic potential of the neutralized system can be
written as \bn\la{c18} \Omega_f&=&\Omega_0-\int
\sigma_1\,d\mu-g\langle A_1^\nu J_\nu^1+A_2^\nu J_\nu^2
\rangle.\en With the goal to find $(\xi_{W^\pm_\nu})^2$, we
calculate $\Omega_f$ near to $T_w$, and we obtain \bn\la{c18}
\Omega_f&=&\Omega_0 -\frac{6\mu^2 T^2}{24}- \langle g'B^0J_0
+gA_3^0J_0^3\rangle
-\frac{\mu^2}{2}\Bigl[(\xi_{W^+_\nu})^2+(\xi_{W^-_\nu})^2\Bigr].
\la{c18.1}\en If we minimize (\re{c18.1}) respect to
$(\xi_{W})^2$, we find that \bn \la{eqtris}
\frac{\partial\Omega_f}{\partial(\xi_{W})^2} &=&
\frac{\partial\Omega_0}{\partial(\xi_{W})^2}-\mu^2=0.\en where we
have taken in consideration that the quantity $\Bigl\langle
g'B^0J_0+ gA^0_3J^3_0\Bigl\rangle$, for $T\approx T_w$, does not
depend over $(\xi_W)^2$. We have also considered that
$(\xi_{W})^2=(\xi_{W^+_\nu})^2=(\xi_{W^-_\nu})^2$ and we have used
$\frac{\partial\xi^2_1}{\partial(\xi_{W})^2}=\frac{g^2}{4\lambda}$.
It is easy to probe that the equation (\re{eqtris}) can be written
as \bn \frac{\partial\Omega_f}{\partial(\xi_{W})^2}
&=&\frac{g^2}{4\lambda}[c^2+\mu^2]\Bigl(1-\frac{T^2}{T^2_c}\Bigr)
+\frac{3g^4}{8\lambda}(\xi_{W})^2-\mu^2=0. \la{c18.3}\en From this
last equation, we can obtain \bn\la{c18.4}
(\xi_{W})^2&=&-\frac{2}{3g^2}\biggl[[c^2+\mu^2]\Bigl(1-\frac{T^2}{T^2_c}\Bigr)
-\frac{4\lambda\mu^2}{g^2}\biggr]
       \no\\
&=&-\frac{2}{3g^2}\biggl[c^2+\mu^2-\Bigl(\frac{g'^2}{4}
+\frac{3g^2}{4}+2\lambda\Bigr)\frac{T^2}{4}
-\frac{4\lambda\mu^2}{g^2}\biggr],\en where we have substituted
$T_c$ by the value given by (\re{ctept}). The critical temperature
of the W boson condensation $T_w$ is obtained from (\re{c18.4})
doing $(\xi_W)^2=0$ and is given by \bn\la{c24}
T^2_w&=&\frac{4(c^2+\mu^2
-\frac{4\lambda\mu^2}{g^2})}{\frac{g'^2}{4}+\frac{3g^2}{4}+2\lambda}.\en
Comparing (\re{c24}) with (\re{c7.1}), we observe that \bn\la{c25}
2\lambda\xi^2_1(T_w)=4\frac{\lambda\mu^2}{g^2},\en and it is
possible to obtain from this last expression the known W
condensation condition given by \ci{PEREZ1,FERRER,KAP1990}
\bn\la{c26} m^2_{W}(\xi_1)=\frac{g^2}{2}\xi^2_1(T_w)=\mu^2,\en
being this fact a good indicator of the theoretical consistency of
the W boson condensation induced by the introduction of the
background charges $j_\mu^1$ and $j_\mu^2$.

If the fermions are included in the ESM, considering the Yukawa
terms in the Lagrangian, the square of the critical temperature of
the electroweak phase transition $T_c$ is given by \be T^2_c =
\frac{4(c^2 + \mu^2)}{\frac{g'^2}{4} + \frac{3g^2}{4} + 2\lambda
+\sum_{m=1}^3[Y^2_{dm} + Y^2_{um} + \frac{Y^2_{em}}{3}]},
\la{cteptwf}\ee being the $Y$'s the Yukawa coupling constants of
the quarks and charged leptons. On the other hand, the square of
the critical temperature of the W boson condensation, for this
case, is \bn\la{ctwgbc} T^2_w&=&\frac{4(c^2+\mu^2
-\frac{4\lambda\mu^2}{g^2})}{\frac{g'^2}{4}+\frac{3g^2}{4}+2\lambda
+\sum_{m=1}^3[Y^2_{dm}+ Y^2_{um} + \frac{Y^2_{em}}{3}]}.\en We
observe that the inclusion of the fermions in the system allows to
low the value of the critical temperature of the W boson
condensation. This fact is also valid for the case of the critical
temperature of the electroweak phase transition.

\section{Conclusions}
In this paper we have shown a W boson condensation in the ESM
induced by the inclusion of background charges in the
thermodynamical system. The W boson condensation that we have
presented here has an origin different from the one studied in
references \ci{PEREZ1}-\ci{KAP1990}. Specifically we have not
included in the system the chemical potentials associated with the
conserved leptonic and neutral weak currents. We have considered
the ESM at finite temperature in presence of only a bosonic
chemical potential $\mu$ associated with the conserved
electromagnetic current. We have neutralized the thermal medium by
the introduction of a background external charge which offsets the
charge density of the scalar field. Particularly we have preserved
the thermodynamic equilibrium of the system by the introduction of
two charges $j_\nu$ and $j^3_\nu$, which are associated with the
$U(1)_Y$ gauge field and the third $SU(2)_L$ gauge field,
respectively.

We have calculated the effective masses of the scalar and gauge
bosons using the mean-field approximation in the high temperature
limit. We have found that the effective mass of the spatial
component of the W boson has a difference of $-\mu^2$ respect to
its temporal component. By the inclusion in the system of the
background charges $j^1_\nu$ and $j^2_\nu$ which are associated
with the first and second gauge fields of the $SU(2)_L$ gauge
group, we have obtained a W boson condensation associated to the
spatial component of W boson. As a consequence of this condensate,
the mentioned difference among the effective masses of the spatial
and temporal components of the W boson has vanished. We have
obtained the critical temperature of the W boson condensation as a
function of $\mu$. We have demonstrated that the W boson
condensate is consistent with the usual condition of condensation
$m^2_W=\mu^2$, where $m_W$ is the effective mass of the W boson.

\section*{Acknowledgments} This work was supported by COLCIENCIAS
(Colombia) under research grant 1101-05-13610.


\begin{thebibliography}{999}
\section*{References}
\bibitem{KAP1981} Kapusta J I (1981) {\it Phys. Rev.} D \textbf{24} 426
\bibitem{WEL1982} Haber H E and Weldon H A (1982) {\it Phys. Rev.} D \textbf{25}
502; (1981) {\it Phys. Rev. Lett.} \textbf{46} 1497
\bibitem{LIN2} Linde A D (1979) {\it Phys. Lett.}  B \textbf{86} 39
\bibitem{KRIVE1980} Krive I V (1980) {\it Yad. Fiz.}  \textbf{31} 1279
[(1980) {\it Sov. J. Nucl. Phys.} \textbf{31} 650]
\bibitem{PEREZ1} P\'erez~Rojas H and Kalashnikov O K (1987) {\it Nucl.
Phys.} B \textbf{293} 241; (1989) {\it Phys. Rev.} D \textbf{40}
1255
\bibitem{PEREZ2} Chaichian M, Montonen C and P\'erez~Rojas H (1991) {\it Phys.
Lett.} B \textbf{256} 227
\bibitem{FERRER} Ferrer E J, De La Incera V and Shabad A E (1987) {\it Phys.
Lett.} B \textbf{185} 407; (1988) {\it Nucl.\ Phys.} B
\textbf{309} 120
\bibitem{KAP1990} Kapusta J I (1990) {\it Phys. Rev.} D \textbf{42} 919
\bibitem{KAPLIB} Kapusta J I (1989) \textit{Finite-Temperature Field Theory}
(Cambridge: Cambridge University Press)
\bibitem{LIN1979} Linde A D (1979) {\it Rep. Prog. Phys.} \textbf{42} 389
\bibitem{SAN2003} Sannino F and Tuominen K (2003) {\it Phys. Rev.} D \textbf{68}
016007
\bibitem{GAV1999} Gavela M B, Pene O, Rius N and Vargas-Castrill\'on S (1999)
{\it Phys. Rev.} D \textbf{59} 025008
\bibitem{FER1987} Ferrer E. J, De La Incera V and Shabad A E (1987)
{\it Nuovo. Cim.} A \textbf{98} 245


\end{thebibliography}
\end{document}